\documentclass[usenatbib]{mn2e}
\usepackage{graphicx,natbib}
\usepackage{amsmath,url}

\newcommand{\aj}{AJ}
\newcommand{\apj}{ApJ}
\newcommand{\apjl}{ApJ}

\newcommand{\mnras}{MNRAS}

\newcommand{\aap}{A\&A}
\newcommand{\aaps}{A\&AS}

\newcommand{\sgra}{Sgr~A$^\star$}

\newcommand{\msun}{{\rm M}_{\odot}}\newcommand{\tmsun}{\mbox{$\msun$}}

\newcommand{\myvec}[1]{{\boldsymbol #1}}
\newcommand\simless{\mathbin{\lower 3pt\hbox
   {$\rlap{\raise 5pt\hbox{$\char'074$}}\mathchar"7218$}}}
\newcommand\simgreat{\mathbin{\lower 3pt\hbox
   {$\rlap{\raise 5pt\hbox{$\char'076$}}\mathchar"7218$}}}

\renewcommand{\labelenumi}{(\alph{enumi})}

\title[Stellar initial mass function in the Galactic Centre]
      {Constraining the initial mass function of stars in the Galactic Centre}
      
\author[U. L\"ockmann, H. Baumgardt, \& P. Kroupa]
{
  U. L\"ockmann\thanks{E-mail: uloeck@astro.uni-bonn.de (UL); holger@astro.uni-bonn.de (HB); pavel@astro.uni-bonn.de (PK)}
  , H. Baumgardt\footnotemark[1], and P. Kroupa\footnotemark[1]\\ 
  Argelander Institute for Astronomy, University of Bonn, Auf dem H\"ugel 71, 53121 Bonn,
  Germany\\
}

\begin{document}

\date{Accepted 2009 October 20. Received 2009 September 12; in original form 2009 July 08}

\pagerange{\pageref{firstpage}--\pageref{lastpage}} \pubyear{2009}

\maketitle

\label{firstpage}

\begin{abstract}
For half a century, evidence has been growing that the formation of stars follows a universal distribution of stellar masses.
In fact, no stellar population has been found showing a systematic deviation from the canonical initial mass function (IMF)
found for example for the stars in the solar neighbourhood.
The only exception may be the young stellar discs in the Galactic Centre, which have been argued to exhibit a top-heavy IMF.

Here we discuss the question whether the extreme circumstances in the centre of the Milky Way may be the reason for a significant variation of the IMF.
By means of stellar evolution models using different codes, we show that the observed luminosity in the central parsec is too high to be explained by a long-standing top-heavy IMF as suggested by other authors, considering the limited amount of mass inferred from stellar kinematics in this region.
In contrast, continuous star formation over the Galaxy's lifetime following a canonical IMF results in a mass-to-light ratio and a total mass of stellar black holes (SBHs) consistent with the observations.
Furthermore, these SBHs migrate towards the centre due to dynamical friction, turning the cusp of visible stars into a core as observed in the Galactic Centre.
For the first time here we explain the luminosity and dynamical mass of the central cluster and both the presence and extent of the observed core, since the number of SBHs expected from a canonical IMF is just enough to make up for the missing luminous mass.

We conclude that observations of the Galactic Centre are well consistent with continuous star formation following the canonical IMF and do not suggest a systematic variation as a result of the region's properties such as high density, metallicity, strong tidal field etc.
If the young stellar discs prove to follow a top-heavy IMF,
the circumstances that led to their formation must be very rare, since these have not affected most of the central cluster.
\end{abstract}


\begin{table*}
  \begin{tabular}{|c|l|c|c|r|r|r|r|r|r|}
    \hline
    Model & IMF & Mass range & $t_0$ & $M_{\rm NS}/M_{\rm tot}$ & $M_{\rm BH}/M_{\rm tot}$ & $N_{\rm NS}$ & $N_{\rm BH}$ & $N_{{\rm mag}_{\rm Ks}<17.5}$ & $M/L_{\rm diffuse}$ \\
    \hline
    (a) & canonical     & $0.01\le m/\msun \le 120$ & -3\,Gyr   & 1.74\,\% & 3.84\,\%  & 16\,861  & 6294     & 11\,442 & 1.86   \\
        &               &                           & $\infty$  & 1.86\,\% & 4.09\,\%  & 18\,064  & 6713     & 7166    & 2.65   \\
        &               &                           & 3\,Gyr    & 1.94\,\% & 4.26\,\%  & 18\,832  & 6989     & 4371    & 3.59   \\
        &               &                           & 1\,Gyr    & 1.97\,\% & 4.31\,\%  & 19\,060  & 7074     & 3736    & 3.95   \\
        &               &                           & 300\,Myr  & 1.97\,\% & 4.32\,\%  & 19\,121  & 7096     & 3630    & 4.03   \\
    \hline
    (b) & $\alpha=1.35$ & $1\le m/\msun \le 120$    & -3\,Gyr   & 8.73\,\% & 67.80\,\% & 80\,853  & 95\,223  & 18\,625 & 2.15   \\
        &               &                           & $\infty$  & 9.31\,\% & 71.98\,\% & 86\,192  & 101\,100 & 8740    & 4.76   \\
        &               &                           & 3\,Gyr    & 9.60\,\% & 74.14\,\% & 88\,879  & 104\,133 & 3152    & 17.91  \\
        &               &                           & 1\,Gyr    & 9.66\,\% & 74.60\,\% & 89\,438  & 104\,781 & 1497    & 74.79  \\
        &               &                           & 300\,Myr  & 9.67\,\% & 74.67\,\% & 89\,519  & 104\,877 & 226     & 942.30 \\
    \hline
    (c) & $\alpha=1.35$ & $0.01\le m/\msun \le 120$ & -3\,Gyr   & 7.27\,\% & 56.43\,\% & 67\,295  & 79\,255  & 14\,107 & 2.44   \\
        &               &                           & $\infty$  & 7.67\,\% & 59.31\,\% & 71\,025  & 83\,309  & 6538    & 4.45   \\
        &               &                           & 3\,Gyr    & 7.87\,\% & 60.78\,\% & 72\,868  & 85\,374  & 2632    & 8.04   \\
        &               &                           & 1\,Gyr    & 7.91\,\% & 61.11\,\% & 73\,262  & 85\,830  & 1966    & 9.61   \\
        &               &                           & 300\,Myr  & 7.92\,\% & 61.18\,\% & 73\,348  & 85\,931  & 1878    & 9.95   \\
    \hline
    (d) & $\alpha=0.85$ & $0.01\le m/\msun \le 120$ & -3\,Gyr   & 5.38\,\% & 85.51\,\% & 48\,567  & 111\,458 & 6236    & 7.45   \\
        &               &                           & $\infty$  & 5.51\,\% & 87.20\,\% & 49\,722  & 113\,673 & 2350    & 16.35  \\
        &               &                           & 3\,Gyr    & 5.56\,\% & 87.89\,\% & 50\,167  & 114\,567 & 687     & 37.29  \\
        &               &                           & 1\,Gyr    & 5.57\,\% & 88.00\,\% & 50\,234  & 114\,712 & 458     & 48.00  \\
        &               &                           & 300\,Myr  & 5.57\,\% & 88.02\,\% & 50\,245  & 114\,738 & 433     & 50.26  \\
    \hline
  \end{tabular}
  \caption
  {Dependence of the composition of the Galactic Centre after 13\,Gyr of star formation on IMF and star formation history.
   The columns give the following in order: Model name; underlying IMF (slope); range of stellar masses used; star formation history parameter; mass fraction in neutron stars; mass fraction in black holes; number of neutron stars and stellar black holes per $1.5\times 10^6\,\msun$ (i.e.\ within 1 pc from \sgra); number of bright stars (${\rm mag}_{\rm Ks}\simless 17.5$); mass-to-light ratio of unresolved population. While most stars survive the age of the Galaxy in the canonical model (a), an old cluster based on a top-heavy IMF is mass-dominated by stellar black holes.
   In contrast to the number of remnants, the mass-to-light ratio as well as the number of bright stars strongly depend on the star formation history. \label{tab:ssemodels}}
\end{table*}

\section{Introduction}

In \citeyear{s55}, Salpeter found that the initial mass distribution of field stars in the range $0.4\simless M_{\star}/\msun \simless 10$ is a power-law with exponent 2.35. Since then, a large number of publications have investigated the initial mass function (IMF) of stars and made clear that star formation in general follows the same empirical law, the canonical IMF \citep[and references therein]{k01}.

Due to its extreme conditions (mass density, velocity dispersion, tidal forces), the Galactic Centre provides a unique environment for testing the universality of the IMF. Star formation in the central region has thus been studied in detail, however no agreement has been reached on the nature of the IMF in either theory or observations: 
\citet{mmt07} find a best fit of observations in the central parsec of our Galaxy with a model of constant star formation with a top-heavy IMF.
On the other hand, \citet{bse09} show that the old stellar cluster in the Galactic Centre very well resembles the bulge population.
Observations of the young, massive Arches cluster in the central region of the Milky Way have long been interpreted as a prime example for top-heavy star formation \citep[e.g.][]{fetal99,setal02,kfkn06,ksj07}. However, \citet{esm09} have shown that a canonical IMF cannot be excluded for this cluster.
\citet{pau06} suggested a flat IMF for the young OB-stars observed in discs in the central parsec from the analysis of the K-band luminosity function.
Based on more recent spectroscopic observations, \citet{bmt+09} find strong evidence for this to be true. 
\citet{br08} found from SPH simulations that the IMF of stars forming in fragmenting accretion discs strongly depends on the parameters of the underlying gas infall scenario.
Unfortunately, theoretical IMF predictions have failed in the past to correctly describe the observations near the Galactic Centre \citep{k08}.

In this paper, we combine observational data with models of stellar evolution and dynamics to constrain the stellar mass function and star formation history in the Galactic Centre.
It is organised as follows: In Section~\ref{sec:ssemodels}, we analyse the properties of models of the Galactic Centre assuming different star formation histories, and compare them to the observations. Section ~\ref{sec:massprof} describes the mass profile of the central parsec and the effect of mass segregation. We discuss the IMF of the young stellar discs around \sgra\ and appropriate formation scenarios in Section~\ref{sec:discimf} and summarise in Section~\ref{sec:conclusions}.

\section{Stellar evolution models of the Galactic Centre}\label{sec:ssemodels}

Observations of the central parsec of the Milky Way show that this region is dominated by a dense population of old stars with a total mass of $\sim 1.5\times 10^6 \msun$ \citep{gso+03,sea+07,sme09,bse09}. Within the uncertainties of a factor of two, \citet{sme09} find that the extended mass inferred from kinematics can be explained well by the visible stars. On the other hand, if star formation in the Galactic Centre occurs following a top-heavy IMF as suggested by \citet{mmt07}, one would expect a large number (and thus significant mass) of dark remnants.

To test which mass functions are consistent with the observations, we used the stellar evolution package \textsc{SSE} \citep{hpt00} to calculate population synthesis models after 13\,Gyr of star formation.
Assuming (broken) power-law IMFs of the form $\xi\left(m\right)\propto m^{-\alpha}$, where $\xi(m) {\rm d}m$ is the number of stars in the mass interval $m$ to $m+{\rm d}m$, we used the following models:

\begin{enumerate}
  \item The canonical IMF according to \citet{k01}, $\xi\left(m\right)\propto m^{-\alpha_i}$, with $\alpha_0=0.3$ $\left(0.01\le m/\msun<0.08\right)$, $\alpha_1=1.3$ $\left(0.08\le m/\msun<0.5\right)$, and $\alpha_2=2.3$ $\left(0.5\le m/\msun \le 120\right)$.
  \item A flat IMF with $\alpha=1.35$ $\left(1\le m/\msun \le 120\right)$ as suggested by \citet{pau06} for the young stellar discs in the Galactic Centre.
  \item The same IMF, but extended to $0.01\le m/\msun \le 120$.
  \item $\alpha=0.85$ as suggested by \citet{mmt07}, again for $0.01\le m/\msun \le 120$.
\end{enumerate}

Starting with solar metallicity ($Z=0.02$), we calculated evolutionary tracks for stars with masses $0.01\le m/\msun \le 120$ in steps of 0.01\,\tmsun.
We averaged the results over time, weighted with different star formation histories of the form ${\rm SFR}(t)\propto e^{-t/t_0}$: We used constant ($t_0=\infty$), exponentially declining ($t_0=3$\,Gyr, $t_0=1$\,Gyr, $t_0=300$\,Myr) and exponentially increasing star formation rates ($t_0=-3$\,Gyr).
Weighing the outcome with any of the IMFs above gives the total mass fraction of neutron stars (NSs) and stellar mass black holes (SBHs), as well as the total numbers of NSs and SBHs in the central parsec (assuming an enclosed mass of $1.5\times 10^6\,\msun$; \citealp{sme09}).
To estimate the total number of bright stars (${\rm mag}_{\rm Ks}\simless 17.5$) and the K-band mass-to-light ratio $M/L_{\rm Ks}$ of the unresolved stars and stellar remnants, we generated a sequence of Padova isochrones \citep{padova_mgb+08,padova_bbc+94}, assuming an average extinction of 3.3\,mag for stars in the Galactic Centre \citep{sme09}. We used the magnitudes calculated for the 2MASS Ks filter \citep{cwm03}, whose transmission curve closely resembles that of the Ks-band filter of the NAOS-CONICA instrument at the ESO VLT, which was used for the observations of the Galactic Centre discussed here.
We further assumed that the mass of a stellar remnant depends on the initial mass as
\begin{equation}
  m_{\rm rem} = \begin{cases}
                      0.109\,m_{\rm init} + 0.394\,\msun, & 0.8 < m_{\rm init}/\msun < 8, \\
                      1.35\,\msun & 8 \le m_{\rm init}/\msun < 25, \\
                      0.1\,m_{\rm init}, & 25 \le m_{\rm init}/\msun.
                  \end{cases}
\end{equation}
\citep{khk+08,tc99,bm08,dkb09}.
The uncertainties of these assumptions are not easily quantified, since the underlying theory and observations are not robust, but
using the masses, radii, and effective temperatures from our \textsc{SSE} models and assuming a blackbody spectrum leads to comparable results.
We are thus confident that our results are correct within a factor much less than two.

\begin{figure}
  \begin{center}
    \includegraphics[width=8.3cm]{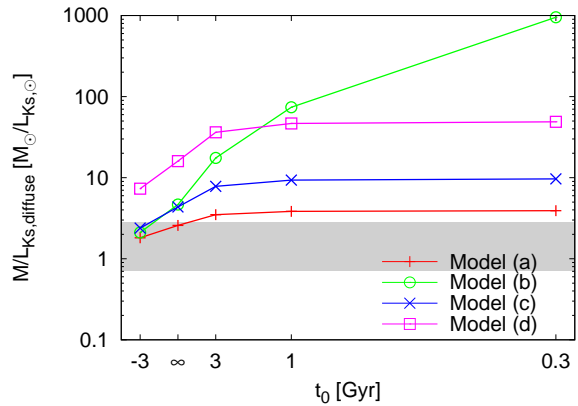}
 \end{center}
  \caption{Ratio of total mass and diffuse light in our models as a function of the star formation history parameter $t_0$.
  The shaded area marks the one-sigma range derived by \citet{sme09} from observations.
  While observations are best explained with an increasing star formation rate following the canonical (a) or a moderately top-heavy (b, c; $\alpha=1.35$) IMF,
  observations cannot be reproduced assuming a flatter IMF (d; $\alpha=0.85$)
  \label{fig:mlratio}}
\end{figure}

Table~\ref{tab:ssemodels} lists the respective values for all four models. While most stars survive the age of the Galaxy as main sequence stars in the canonical model (a), an old cluster based on a top-heavy IMF is mass-dominated by stellar black holes.
\citet{sme09} find that the unresolved stellar population makes up $>98\%$ of the mass in the central parsec, and find its mass-to-light ratio to be $M/L_{\rm Ks} = 1.4^{+1.4}_{-0.7}\,\msun/L_{\odot,\rm Ks}$.
Figure~\ref{fig:mlratio} plots the ratio of total mass to diffuse light in our models as a function of the star formation history parameter $t_0$.
We find that the observations are consistent with the canonical IMF (a), with a tendency towards constant or increasing star formation.
The $\alpha=1.35$ models (b, c) require increasing star formation, and an IMF as flat as $\alpha=0.85$ (d) is not consistent with the observed old population at all.

We can also compare the results of our analysis as given in Table~\ref{tab:ssemodels} to stellar number counts in the central parsec. Assuming a canonical IMF, we expect $\sim 2.5 \times 10^4$ SBHs and NSs for every $1.5 \times 10^6\,\msun$ in stars and stellar remnants, the latter being the estimated enclosed mass within 1\,pc from \sgra. Due to dynamical friction, SBHs may migrate to the central parsec from as far as 6\,pc from the centre within 10\,Gyr (see Section~\ref{sec:massprof} for details), thus increasing the number of SBHs in the central parsec by up to one order of magnitude. On the other hand, SBHs may spiral into the SMBH as discussed above. In total, the expected number of SBHs agrees best with \citet{mpb+05} suggesting a number of $\sim 10^4$ SBHs and NSs in this region from X-ray observations, while models based on a top-heavy IMF suggest numbers well above $10^5$.

A more verifiable quantity is the number of bright stars: \citet{sme09} find $\sim 6000$ stars with Ks-band magnitude $\simless 17.5$ within a projected distance from \sgra\ of 1\,pc. Assuming a spherical distribution with a density profile $\rho \propto r^{-1.75}$ \citep{sea+07}, $\sim 3000$ of these stars are in the innermost parsec. This value is consistent with our model (a) of a canonical IMF, assuming a constant or declining star formation rate, or with models (b) and (c) assuming constant or slightly decreasing star formation ($t_0 > 3$\,Gyr). An even flatter IMF (d) requires a star formation rate increasing with time to explain the observed number of luminous stars.

\begin{figure*}
  \begin{center}
    \includegraphics[width=16.6cm]{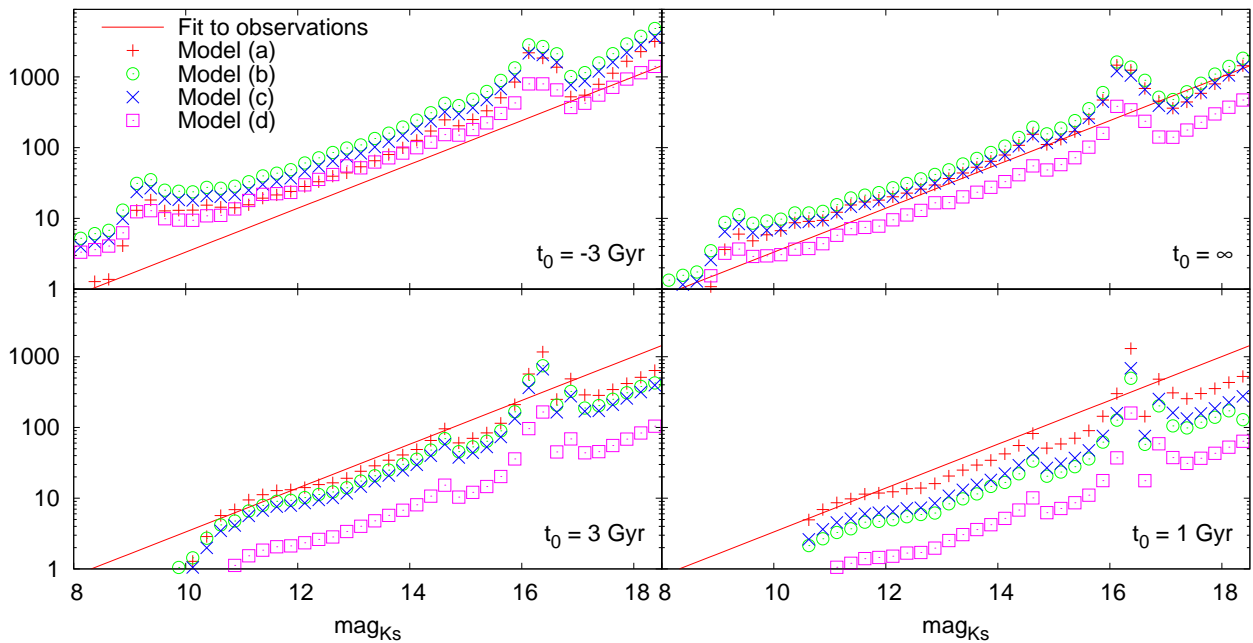}
 \end{center}
  \caption{Luminosity function of bright stars in the Galactic Centre, assuming a K-band extinction of 3.3\,mag.
    The panels compare the LF observed by \citet{bse09}, represented by their fit as a straight line (and corrected for projection effects), to the results of our models of different IMFs following exponentially increasing (upper left), constant (upper right), and exponentially decreasing star formation (lower panels).
    The slope derived from the observations fits all our models quite well: It does not depend strongly on star formation history or IMF.
    However, as is seen also from Table~\ref{tab:ssemodels}, the normalisation does depend on the star formation rate, especially so for the top-heavy models (d, b).
    \label{fig:klf}}
\end{figure*}

\citet{bse09} find that the K-band luminosity function (LF) of late-type stars in the central parsec closely resembles that of the bulge population. Here we use the luminosities calculated from our stellar evolution models to compare the LFs of different IMFs. Figure \ref{fig:klf} shows the K-band LFs of our models as a function of star formation history. It is seen that the shape of the LF does not depend significantly on the IMF. In particular, all curves can be approximated by a power-law with a slope $\beta \approx 0.3$, and all exhibit the horizontal branch / red clump peak.
This peak is offset to the observations of \citet{bse09} by one magnitude, which may be due to our models of stellar evolution and the assumed constant extinction. We thus presume an uncertainty of a factor of two in the luminosity estimates discussed above, which does not affect our qualitative results and reasoning.

Altogether, observations of the old stars, diffuse light, and stellar dynamics in the Galactic Centre are best explained with a canonical IMF at constant or decreasing star formation, but may also be explained with a somewhat flatter IMF and an almost constant star formation rate.
An IMF slope flatter than $\alpha = 1$ can be safely ruled out.

We have to stress that here (and in the next section) we only calculate self-contained models, i.e.\ we assume that the stars now present in the Galactic Centre also formed in this region. Stars may also be brought to the vicinity of the SMBH by capture of individual stars or clusters, and some unknown fraction of the stars observed in the centre of the Milky Way may have formed in a much different environment.
To this extent, we do not exactly discuss star formation in the Galactic Centre, but the formation of stars now observed in the central region, thus reaching an insight into the composition of stellar systems surrounding SMBHs.

\section{Mass profile in the Galactic Centre}\label{sec:massprof}

Theoretical arguments and $N$-body simulations show that a stellar system around a SMBH evolves into a cusp with a $\gamma = 1.75$ power-law density distribution \citep{bw76,bme04a,bme04b,pms04,afs04}.
Indeed, observations show that the central parsec of the Milky Way exhibits a corresponding profile \citep[e.g.][]{gs03,sea+07,sme09}.
However, observations also reveal a deficit of stars within a few $0.1$\,pc from the SMBH \citep{gtk+96,gso+03,fgk+03,sea+07,sme09}:
For example, \citet{sea+07} estimated the density profile of the stellar cusp in the Galactic Centre from the observed luminosity profile and kinematics as
\begin{equation}
\rho(r)=\left(2.8 \pm 1.3\right)\times 10^6\,\msun {\rm pc}^{-3} \left(\frac{r}{0.22\,{\rm pc}}\right)^{-\gamma}, \label{eq:rho}
\end{equation}
where $r$ is the distance from the SMBH, $\gamma=1.2$ inside 0.22\,pc and $\gamma=1.75$ outside 0.22\,pc.
\citet{bse09} find that the profile may be flatter or even slightly inverted in the central 0.2\,pc, turning the cusp into a core.

The observed core in the old population around \sgra\ may be explained by stellar collisions destroying the envelopes of giants \citep{gtk+96,detal98,a99,bd99,ddcf09}, or by an inspiralling IMBH depleting the central region \citep{bgp06,lb08,ms06}.

Here we suggest that the central 0.2\,pc around \sgra\ are not mass-depleted, but dominated by a cusp of SBHs having displaced the less massive visible stars:
Due to dynamical friction, the most massive stellar remnants sink into the depth of the SMBH's potential well. Thus, the SMBH is expected to be surrounded by a large number of black holes in its immediate vicinity \citep{bme04b,fak06}. \citet{fak06} have shown that the SBHs in the Galactic Centre establish a cusp profile $\rho(r) \propto r^{-1.75}$, while the density profile of the less massive stars becomes consistent with the observed luminosity profile. They do not find a clear density cutoff around 0.2\,pc as in the observations, since they had to use an unrealistically large number of SBHs due to computational limitations of the methods used. Since it is as yet not possible to perform a full calculation including a realistic number of stars and SBHs, we here follow a theoretical approach.

For simplicity, we assume that the Galactic Centre formed evolved stars and SBHs 10\,Gyr ago without any further star formation.
The frictional drag on an inspiralling SBH can be estimated as
\begin{equation}
 \frac{{\rm d}\myvec{v}}{{\rm d}t} = -\frac{4 \pi \ln{\Lambda} G^2 \rho_{\star}(r,t) M_{\rm BH}}{v^3}
\left[ {\rm erf}(X) - \frac{2 X}{\sqrt{\pi}} e^{-X^2} \right] \myvec{v}
 \label{dynf}
\end{equation}
(\citealp{bt87}, Eq.~7-18), where $\myvec{v}$ is the velocity of the SBH, $G$ is the gravitational constant, $M_{\rm BH}$ is the SBH mass,
$\rho_{\star}(r,t)$ is the density of the stellar background,
$\ln{\Lambda}$ is the Coulomb logarithm and $X=v/(\sqrt{2}\sigma)$
is the ratio between the SBH velocity and the (1D) stellar velocity dispersion $\sigma$.
For a stellar density profile $\rho_{\star}(r)\propto r^{-\gamma}$ with $1.2 \leq \gamma \leq 1.75$, one gets $X \approx 1.2$.
Assuming that the overall density profile does not change with time and can be described by $\rho(r) \equiv \rho_{\star}(r,t) + \rho_{\rm BH}(r,t) = \rho_0 r^{-\gamma}$ yields
\begin{equation}
  v_c(r) = \sqrt{G\left(M_{\rm SMBH} + \frac{4 \pi \rho_0}{3-\gamma} r^{3-\gamma}\right) r^{-1}}
\end{equation}
as the circular velocity at distance $r$ of the SMBH of mass $M_{\rm SMBH}$. Inserting into
\begin{equation}
    -r \left|\frac{{\rm d}\myvec{v}}{{\rm d}t} \right|
  = -\frac{F r}{M_{\rm BH}} = \frac{{\rm d}L}{{\rm d}t}
  = \frac{{\rm d}}{{\rm d}t} \sqrt{r v_c},
\end{equation}
where $F$ and $L$ are the frictional force and orbital angular momentum, respectively, leads to the SBH inspiral speed
\begin{equation}
    \dot{r}
  = \frac{-8 \pi r^{5/2}\ln{\Lambda} \sqrt{G} \rho_{\star}(r,t) M_{\rm BH} \left[ {\rm erf}(X) - \frac{2 X}{\sqrt{\pi}} e^{-X^2} \right]}
    {\left(M_{\rm SMBH} + 4 \pi \rho_0\frac{4-\gamma}{3-\gamma} r^{3-\gamma}\right)
         \sqrt{M_{\rm SMBH} + \frac{4 \pi \rho_0}{3-\gamma} r^{3-\gamma}}}
\end{equation}
assuming a circular orbit of the SBH. Note that $\rho_{\star}(r,t)$ decreases with time due to the inspiralling SBHs increasing $\rho_{\rm BH}(r,t)$.

\begin{figure}
  \begin{center}
    \includegraphics[width=8.3cm]{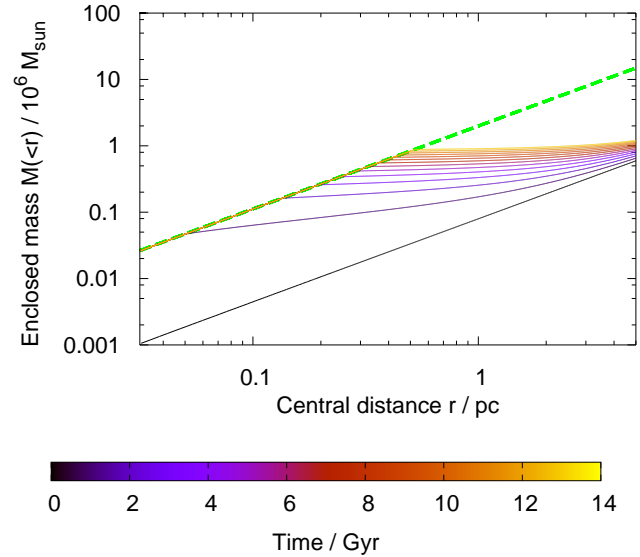}
 \end{center}
  \caption{Mass segregation in the Galactic Centre. The solid lines show the enclosed mass in SBHs as a function of central distance in steps of 1\,Gyr from the black bottom curve ($t=0$) to the bright top curve ($t=13$\,Gyr). The dashed line depicts the total enclosed mass in SBHs and stars, which is assumed to be constant. It can be seen that the central 0.5\,pc are saturated with SBHs over the Galaxy's lifetime.
    \label{fig:sbh-inspiral}}
\end{figure}

To integrate this equation numerically, we assume 
that the cluster was initially not
mass-segregated, $\rho_{\rm BH}(r,0) = 0.04\,\rho(r)$, where $M_{\rm BH}/M_{\rm tot} \approx 0.04$ was taken from Table~\ref{tab:ssemodels} for a canonical IMF, and $\rho(r) = 2.8\times 10^6\,\msun {\rm pc}^{-3} \left(r/0.22\,{\rm pc}\right)^{-\gamma}$ is a \citet{bw76} profile with $\gamma=1.75$. We further assume a SMBH mass of $M_{\rm SMBH}=4\times 10^6\,\msun$, and SBHs of mass $M_{\rm BH} = 10\,\msun$.
Figure~\ref{fig:sbh-inspiral} shows the enclosed mass in SBHs as a function of central distance, $M_{\rm SBH(<r)}$, for different times $t$ in steps of 1\,Gyr, starting with $M_{\rm SBH(<r)} \propto r^{3-\gamma}$ at $t=0$. It is seen that the cusp is saturated by SBHs in the innermost 0.5\,pc within 10\,Gyr. Clearly, this is not the final answer: Due to random deflections of stars, we can expect a number of stars in the innermost region. Furthermore, contrary to our assumptions above, both stars and SBHs move on eccentric orbits, preventing a strict segregation.

In fact, \citet{fak06} find that the inspiralling SBHs build a \citet{bw76} profile, while the density profile of the stars is flatter, compatible with the slope observed in the central 0.2\,pc. In particular, they find that the central part will be mass-dominated by the SBHs, suggesting that the estimates above are at least qualitatively correct.

Altogether, the above discussion suggests that
\begin{itemize}
  \item the innermost region is mass-dominated by SBHs within $a \sim 0.5$\,pc, the semi-major axes $a$ being distributed as $a^{2-1.75}$,
  \item in this region, the stars are distributed roughly as $a^{2-1.2}$, and
  \item outside $a \sim 0.5$\,pc, the cluster is strongly dominated by visible stars, $\rho_{\star} \simgreat 0.94 \rho$ assuming a canonical IMF.
\end{itemize}
To compare these results to the density profile derived from observations, we need to convert the distributions of semi-major axes into a distance distribution. For this, we assume a thermal distribution (i.e.\ a uniform distribution in $e^2$, where the eccentricity $e$ is distributed as $f_e(e) = 2e$). A cusp with 3D ``density'' of semi-major axes $\rho_a(a)$ has a 1D distribution $\phi_a(a) = 4 \pi a^2 \rho_a(a)$.
The central distance of a particle on a Kepler orbit with semi-major axis $a$ and eccentricity $e$ passes any value $r$ (with $1-e<r/a<1+e$) twice during a full orbital period $T$, hence its distribution function is
\begin{equation}
f_r(r)= \frac{2}{T}\left|\frac{{\rm d}r}{{\rm d}t}\right|^{-1} = \frac{r}{\pi a \sqrt{a^2e^2 - (a-r)^2}}.
\end{equation}
A particle on an orbit of eccentricity $e$ and semi-major axis $a$ assumes central distances $r$ between pericentre and apocentre distance, $a(1-e)\le r \le a(1+e)$.
Depending on the orbital eccentricity, a particle at distance $r$ can thus have a semi-major axis between $r/2$ and infinity. For a given semi-major axis $a$, the eccentricity has to be larger than $\left|1-r/a\right|$ to be consistent with the central distance $r$.
Altogether, the distribution function of central distances is
\begin{eqnarray}
  \phi_r(r) &=& \int_{a=r/2}^{\infty} \int_{e=\left|1-r/a\right|}^1 f_e(e) \phi_a(a) f_r(r) \,{\rm d}e\,{\rm d}a\\
            &=& \int_{a=r/2}^{\infty} \int_{e=\left|1-r/a\right|}^1 \frac{8 a\, r\, e\, \rho_a(a)}{\sqrt{a^2e^2 - (a-r)^2}} \,{\rm d}e\,{\rm d}a,
\end{eqnarray}
and with $\phi_r(r) = 4 \pi r^2 \rho(r)$, the 3D density profile is
\begin{eqnarray}
  \rho(r) = \int_{a=r/2}^{\infty} \int_{e=\left|1-r/a\right|}^1 \frac{2 a\, e\, \rho_a(a)}{\pi r \sqrt{a^2e^2 - (a-r)^2}} \,{\rm d}e\,{\rm d}a.
\end{eqnarray}

\begin{figure}
  \begin{center}
    \includegraphics[width=8.3cm]{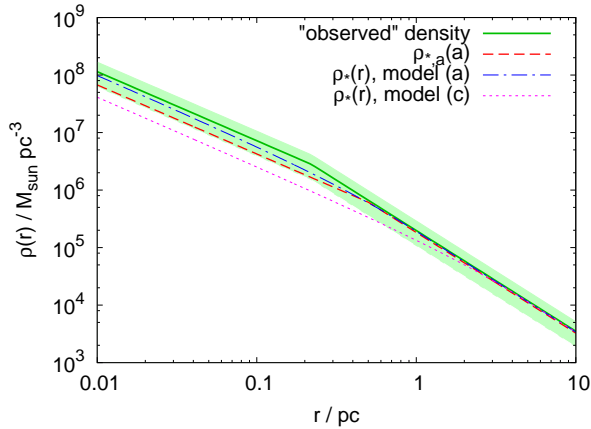}
 \end{center}
  \caption{Density and semi-major axis distributions of stars in the Galactic Centre. The dashed line shows a semi-major axis distribution $\rho_{\star,a}(a)$ with a break radius of 0.5\,pc, as predicted by theory for a canonical IMF. Assuming a thermal eccentricity distribution, the dash-dotted line shows the corresponding density profile $\rho_{\star}(r)$, which resembles the profile derived from observations by \citet{sea+07} very well. Both distributions are consistent with the observed profile within the $1\sigma$ uncertainty shown as a shaded area. In contrast, the expected density profile for a top-heavy IMF (model (c) in Table~\ref{tab:ssemodels}; dotted line) is not consistent with the observations, producing a break radius at 2.5\,pc already after 8\,Gyr of mass segregation.
    \label{fig:dens}}
\end{figure}

Figure~\ref{fig:dens} shows a semi-major axis distribution $\rho_{\star,a}(a) = 2.8 \times 10^6\,\msun {\rm pc}^{-3}  (0.5/0.22)^{-\gamma} \times 0.96$ of stars in the Galactic Centre, as well as the resulting 3D density profile $\rho_{\star}(r)$. It can be seen that the density profile derived in our theory resembles the observations very well. The exact value of the break radius cannot be determined within the uncertainties of the density profile derived from the observations by \citet{sea+07}.

On the other hand, models with a black hole mass ratio $M_{\rm BH} / M_{\rm tot} \simgreat 0.6$, as would result from the evolution of a population of stars following a top-heavy IMF (cf.\ Table~\ref{tab:ssemodels}), lead to a break radius of 2.5\,pc and higher already after 8\,Gyr of mass segregation. Thus, the observed break radius is further evidence for a canonical IMF and a moderate black hole fraction in the Galactic Centre.

\section{IMF of the young stellar discs}\label{sec:discimf}

Observations of the Galactic Centre revealed one or two discs of $\approx 6$\,Myr old stars orbiting the central super-massive black hole (SMBH)
at a distance of $\sim 0.1$\,pc (\citealp{lb03,pau06,lu06,lu09,bmf08}; see discussion in \citealp{lb09}).
Two main scenarios have been proposed for the formation of such massive ($M_{\rm disc} \sim 10^4\,\msun$) discs close to \sgra:
\citet{ger01} suggested that a disc of stars may have formed by tidal disruption of an infalling cluster of young stars, which would require a large mass or a central intermediate-mass black hole to survive the strong tidal forces from the SMBH \citep{mp03}.
However, \citet{lb03} showed that this cannot explain the small distances of the stars from \sgra, and propose in-situ formation by fragmentation of a massive accretion disc as an alternative scenario \citep[see also][]{nc05}.
\citet{mhmw08} and \cite{br08} have shown the infall of a giant molecular cloud (GMC) towards the Galactic Centre to be effective in creating a disc of stars with distances from \sgra\  consistent with the observed young massive stars.

\citet{pau06} derive a flat IMF for the observed disc stars from their K-band luminosity function; however, this only refers to a mass interval of $20\simless M_{\star}/\msun \simless 30$. 
From the amount of mutual warping of the observed stellar discs, \citet{ndcg06} derived upper mass limits to the discs' masses considering their apparent flatness as measured by \citet{pau06} after a few Myr of interaction.
However, in \citet{lbk09} we show that a canonical IMF cannot be excluded from disc dynamics: \citet{ndcg06} used stellar discs only out to 0.2\,pc from the SMBH, while more than one third of the stars listed by \citet{pau06} are further away from \sgra.
In addition, they did not distinguish between stars in the outer parts of the discs which almost retain their orbital planes, and stars close to the centre which strongly precess.

\citet{ns05} argue that the X-ray luminosity of the \sgra\ field is
 too low to account for the number of young $\simless\, 3\,\msun$ stars expected from a canonical IMF, considering the large number of O-stars observed in the discs, which
may be explained by a higher low-mass cutoff near 1\,\tmsun.
On the other hand, the existence of the S-stars within 0.01\,pc from \sgra\ is in favour of a canonical IMF (and thus a large number of B-type stars) for the stellar discs, if they were formed from these discs as suggested by \citet{lbk08}.
For a more detailed discussion of the IMF of the young stellar discs, see \citet{lbk09}.

Assuming an infalling cluster as the origin for the stellar discs, one may expect a top-heavy mass function if the cluster was mass segregated and then tidally stripped. On the other hand, a top-heavy \emph{initial} mass function of stars formed in a fragmenting disc would require an unusual mode of star formation. As we have discussed in the previous sections, no convincing reason or evidence for a flat IMF in the Galactic Centre in general has been found.

Using SPH simulations of star formation in fragmenting gas accretion discs, \citet{br08} find that the IMF of disc stars can be bimodal (and thus top-heavy) if the infalling gas cloud is massive enough ($\simgreat 10^5\,\msun$) and the impact parameter of the SMBH encounter is as small as $\sim 0.1\,$pc.
This way, an extreme configuration of cloud masses and distances may lead to a significant variation of the IMF.

Only recently, a systematic search of OB stars in the central parsec revealed a significant deficit of B-type stars in the regime of the young discs, suggesting a strongly top-heavy IMF for these discs \citep{bmt+09}.
Until then, there was no strong evidence for a top-heavy IMF in the young discs.
However, it cannot be assumed that the majority of stars in the Galactic Centre were formed the same way as the young disc stars, whose existence may be an indication of recently enhanced star formation processes: If the Galactic Bar is young, as an increased fraction of barred galaxies for lower redshifts suggests \citep{see+08}, bar-induced gas inflow may explain such an enhancement by an increasing supply of high-mass GMCs towards the central region (\citealp{sw93} and references therein). Hence, despite
the significant lack of B-stars in the range $0.03-0.5$\,pc from the SMBH, suggesting that the young discs indeed formed following a flat IMF, this does not imply that star formation in the Galactic Centre is or was in general top-heavy.
Instead, the mass function of disc stars may reveal details of the formation scenario, as the results of \citet{br08} suggest.
The majority of old stars in the region may thus have formed under different conditions (e.g.\ mass and impact parameter of the infalling clouds) in the scenario of a fragmenting disc, or formed further away and then migrated to the centre (as in the infalling cluster scenario, see above), or formed by any other process following the canonical IMF.

\section{Discussion}\label{sec:conclusions}

Various attempts have been made to study star formation in the Galactic Centre, but so far neither theory or simulations nor observations led to an agreement on the (initial) distribution of stellar masses. Here we have shown that theory and observations are consistent with star formation generally following a canonical IMF \citep{k01} in the Galactic Centre, just as anywhere else in the Universe.
Our main results can be summarised as follows:
\renewcommand{\labelenumi}{\arabic{enumi}.}
\begin{enumerate}
  \item The mass-to-light ratio of the central parsec of the Milky Way is consistent with a constant or exponentially decreasing star formation rate following a canonical IMF. Models of constant star formation following an IMF with $\alpha = 1.35$ are consistent with the observed luminosities but create $\sim 10^5$ SBHs in the central parsec, ten times more than expected by other authors.
  Mass functions flatter than $\alpha \approx 1$
   can be safely ruled out, since they cannot explain the observed number of bright stars and the diffuse light.
  \item The core observed in the luminosity distribution with a radius of $r_{\rm break} \approx 0.2$\,pc does not imply a core in the mass profile, but can be well explained by mass segregation as suggested by \citet{fak06}, where dark remnants mass-dominate this region.
  Again, the observations are best explained by a canonical IMF, and are not compatible with star formation following a top-heavy IMF with $\alpha \le 1.35$, as this would create a core radius one order of magnitude larger.
  \item Recent observations revealing a deficit of B-type stars in the young stellar discs suggest a top-heavy IMF for this population, which may be explained by tidal stripping of an infalling mass-segregated cluster, or unusual modes of star formation in a fragmenting accretion disc.
 However, these results do not allow conclusions on star formation in the Galactic Centre in general, for which we have no reason to assume it to be non-canonical.
\end{enumerate}

While other authors generally predicted a flat IMF \citep[e.g.][]{ksj07} or a higher low-mass cut-off \citep[e.g.][]{m93,lb03,l_06} from state-of-the-art theoretical star formation models of the Galactic Centre, we find that observations suggest star formation follows a canonical IMF even under the extreme circumstances present in the central cluster. This universality of the IMF poses a major challenge to our understanding of star formation processes (see also \citealp{k01,k08b}).

Our results rely on the assumption that the stars observed within 1\,pc from the Galactic Centre also formed there, suggesting star formation with a canonical IMF even under such exotic conditions. It is possible that some of the stars were brought in by massive star clusters which spiralled towards the SMBH through dynamical friction \citep{pbmmhe06, fifm08}. However, this scenario is unlikely because a star cluster is stripped on its way towards the centre and loses mostly low-mass stars, since it would be in a mass segregated state soon after its formation. Therefore, the most likely scenario is a central cluster that formed over a Hubble time with a canonical IMF, where the very young stellar population of the stellar discs observed to have a very top-heavy IMF \citep{bmt+09} constitutes a rare star formation event not typical for the bulk stellar population in the central cluster.

\section*{Acknowledgments}
This work was supported by the German Research Foundation (DFG) through the priority programme 1177
`Witnesses of Cosmic History: Formation and Evolution of Black Holes, Galaxies and Their Environment'.

\bibliographystyle{aa_mn2e}

\makeatletter   \renewcommand{\@biblabel}[1]{[#1]}   \makeatother

\label{lastpage}

\end{document}